\documentclass[aps,prl,floatfix,twocolumn,showpacs,superscriptaddress]{revtex4}
\usepackage{bm}
\usepackage{epsf}
\usepackage{float}
\usepackage{amssymb}
\usepackage{amsmath}
\usepackage{graphicx}
\usepackage{epsfig}
\usepackage{amsmath}
\usepackage{color}

\newcommand*{\Scale}[2][4]{\scalebox{#1}{$#2$}}%
\def\scalefactor{1.5}

\begin{document}

\title{Spontaneous quantum Hall effect in an atomic spinor Bose-Fermi mixture}
\author{Zhi-Fang Xu}
\affiliation{Department of Physics and Astronomy, University of Pittsburgh,
  Pittsburgh, Pennsylvania 15260, USA}
\author{Xiaopeng Li}
\affiliation{Condensed Matter Theory Center and Joint Quantum Institute,
Department of Physics, University of Maryland, College Park, MD 20742-4111, USA}
\author{Peter Zoller}
\affiliation{Institute for Quantum Optics and Quantum Information of the Austrian Academy of Sciences, A-6020 Innsbruck, Austria}
\affiliation{Institute for Theoretical Physics, University of Innsbruck, A-6020 Innsbruck, Austria}
\author{W. Vincent Liu}
\affiliation{Department of Physics and Astronomy, University of Pittsburgh,
  Pittsburgh, Pennsylvania 15260, USA}
\affiliation{Wilczek Quantum Center, Zhejiang University of Technology, Hangzhou 310023, China}

\begin{abstract}
 We study a mixture of spin-$1$ bosonic and spin-$1/2$ fermionic cold atoms, e.g., $^{87}$Rb and $^{6}$Li, confined in a triangular optical lattice. With fermions at $3/4$ filling, Fermi surface nesting leads to spontaneous formation of various spin textures of bosons in the ground state, such as collinear, coplanar and even non-coplanar spin orders. The phase diagram is mapped out with varying boson tunneling and Bose-Fermi
 interactions. Most significantly, in one non-coplanar state the mixture is found to exhibit a spontaneous quantum Hall effect in fermions and  crystalline superfluidity in bosons, both driven by interaction.
\end{abstract}

\date{\today}


\maketitle

{\it Introduction.}
Searching for topological states of quantum matter, such as quantum Hall states and topological insulators, is one of the most important directions in ultracold atoms. To achieve such novel states in a charge neutral system, external manipulations to synthesize effective gauge fields ~\cite{Dalibard2011} have been carried out, for example, by rotating the system~\cite{Cooper2008}, using Raman laser field coupling schemes ~\cite{Aidelsburger2011,Jimez-Garc2012,Aidelsburger2013,Miyake2013,Aidelsburger2014,Lin2011,Zhang2012,Wang2012,Cheuk2012}, or shaking the lattice~\cite{Struck2011,Struck2012,Hauke2012,Parker2013}.
At present the realization of topological states has remained an experimental challenge. An alternative route
is to have fermions move in a spontaneous spin texture background, which can be stabilized by Ruderman-Kittel-Kasuya-Yosida (RKKY) interactions in classical Kondo lattice  models (CKLM)~\cite{Ohgushi2000,Taguchi2001,Shindou2001,Martin2008,Akagi2010,Akagi2012}.
For example, a triple-$\mathbf{Q}$ magnetic ordering is favored in a triangular lattice at $3/4$ Fermi filling and it leads to a quantum Hall effect~\cite{Martin2008,Akagi2010}. However, what atomic systems can support such mechanism is an open question.

In this Letter, we show that topologically nontrivial spin textures and a quantum Hall effect can be naturally realized in an atomic mixture of spin-$1$ bosons and spin-$1/2$ fermions, e.g., $^{87}$Rb and $^{6}$Li,
loaded into a two-dimensional triangular optical lattice.
Spin-1 bosonic $^{87}$Rb atoms with a ferromagnetic interaction ~\cite{Ho1998,Ohmi1998} provide large local moments, i.e., effective classical spins, when they are condensed. Interspecies spin-changing collisions between Rb and Li atoms give rise to  a spin-exchange interaction, which plays the role of a Kondo coupling. Performing a self-consistent mean field study for the ground state of the system at $3/4$ Fermi filling, we find various topologically distinct spin textures such as collinear, coplanar and  non-coplanar.
While the RKKY mechanism is thus confirmed to  play an important role, the spinor Bose-Fermi mixture model actually contains important ingredients beyond the description of CKLM. Two key ones are the number fluctuation of bosons and Bose-Fermi density interactions, which are responsible for the following richer quantum phenomena beyond the scope of CKLM. Due to effective gauge fields in the non-coplanar state, bosons are predicted to condense at a finite momentum leading to chiral superfluidity.
The novel chirality of bosons is not only interesting by itself but also provides experimental fingerprints for the emergence of gauge fields in time-of-flight measurements. Moreover, several spin texture and even ferromagnetic states are accompanied by density wave (DW) orders which are attributed to Bose-Fermi density interactions~\cite{Orth2009}.

{\it Spinor Bose-Fermi mixture.}
Consider a mixture of spin-1 bosonic and spin-1/2 fermionic cold atoms.
There will be the usual density and spin interactions between two particles in the same species~\cite{Stamper-Kurn2013,Kawaguchi2012}.
The crucial new ingredient is the interaction between the species of bosons and fermions, which is described by the contact pseudo-potential
$V_{bf}(\mathbf{r}_1-\mathbf{r}_2)=\left(g_{1/2}\hat{\mathcal{P}}_{1/2} +g_{3/2}\hat{\mathcal{P}}_{3/2}\right)\delta(\mathbf{r}_1-\mathbf{r}_2)$,
where $g_{F_{\rm tot}}=2\pi\hbar^2 a_{F_{\rm tot}}/m_{bf}$, $a_{F_{\rm tot}}$ is the s-wave scattering length in the channel of total spin $F_{\rm tot}$, $\hat{ \mathcal{P}} _{F_{\rm tot}}$ is the corresponding projection operator, and $m_{bf}$ is the reduced mass. Based on the identity $2\hat{\mathbf{S}}\cdot\hat{\mathbf{F}}=\hat{\mathcal{P}}_{3/2}-2\hat{\mathcal{P}}_{1/2}$ \cite{Stamper-Kurn2013},
where $\hat{\mathbf{S}}$ and $\hat{\mathbf{F}}$ are the one-particle vector spin for spin-$1/2$ and spin-$1$ atoms, respectively.
The Bose-Fermi contact interaction can be rewritten as
\begin{eqnarray}
V_{bf}(\mathbf{r}_1-\mathbf{r}_2)&=\left(g_d \hat{I}_S\otimes\hat{I}_F+g_s \hat{ \mathbf{S}}  \cdot \hat{\mathbf{F}} \right)\delta(\mathbf{r}_1-\mathbf{r}_2),
\label{interaction2}
\end{eqnarray}
where $g_d=(g_{1/2}+2g_{3/2})/3$ and $g_s=(2g_{3/2}-2g_{1/2})/3$ are the density-density interaction and spin-exchange interaction strengths, respectively. Here, $\hat{I}_S$ ($\hat{I}_F$) is the identity operator for the fermion (boson).

With the spinor Bose-Fermi mixture loaded into  a triangular optical lattice~\cite{Becker2010}, the system is approximately described  by a tight-binding model Hamiltonian $\hat{H}=\hat{H}_b+\hat{H}_f+\hat{H}_{bf}$, where
\begin{eqnarray}
\hat{H}_f&=&-t_f\sum\limits_{\langle i,j\rangle,\sigma}\hat{c}_{i,\sigma}^{\dagger}\hat{c}_{j,\sigma}
+U_f\sum\limits_{i}\hat{n}_{f,i,\uparrow}\hat{n}_{f,i,\downarrow},\nonumber\\
\hat{H}_b&=&-t_b\sum\limits_{\langle i,j\rangle,\alpha}\hat{b}_{i,\alpha}^{\dagger}\hat{b}_{j,\alpha} +\frac{U_b}{2}\sum\limits_{i} \hat{n}_{b,i}(\hat{n}_{b,i}-1)
\nonumber\\
&&+\frac{J_b}{2}\sum\limits_{i} (\hat{\mathbf{F}}_{i}^2-2\hat{n}_{b,i}),\nonumber\\
\hat{H}_{bf}&=&U_{bf}\sum\limits_{i}\hat{n}_{f,i}\hat{n}_{b,i}
+J_{bf}\sum\limits_{i} \hat{\mathbf{S}}_{i}\cdot\hat{\mathbf{F}}_{i}.
\label{hamiltonian}
\end{eqnarray}
Here, $\langle i,j\rangle$ denotes the summation over nearest-neighbor sites and $\sigma=\uparrow,\downarrow$ and $\alpha=+1,0,-1$ denote spin states of fermionic and bosonic atoms, respectively. $\hat{c}_{i\sigma}$ ($\hat{a}_{i\alpha}$) is the annihilation operator for fermions (bosons) with spin $\sigma$ ($\alpha$). $\hat{n}_{f,i}=\sum_{\sigma}\hat{n}_{f,i,\sigma}=\sum_{\sigma}\hat{c}_{i,\sigma}^{\dagger}\hat{c}_{i,\sigma}$ ($\hat{n}_{b,i}=\sum_{\alpha}\hat{n}_{b,i,\alpha}=\sum_{\alpha}\hat{b}_{i,\alpha}^{\dagger}\hat{b}_{i,\alpha}$) is the number of fermions (bosons) at site $i$. The spin operator $\hat{\mathbf{S}}_i$ ($\hat{\mathbf{F}}_i$) has three components $\hat{S}_{i\nu}=\sum_{\sigma\sigma'}\hat{c}_{i,\sigma}^{\dagger}(S_{\nu})_{\sigma\sigma'}\hat{c}_{i,\sigma'}$ ($\hat{F}_{i\nu}=\sum_{\alpha\alpha'}\hat{b}_{i,\alpha}^{\dagger}(F_{\nu})_{\alpha\alpha'}\hat{b}_{i,\alpha'}$) with $\nu=x,y,z$, where $S_{\nu}$ ($F_{\nu}$) is the $\nu$ component of the spin-$1/2$ (spin-$1$) operator.
$U_f$, $U_b$, and $U_{bf}$ represent on-site density interaction strengths, and $J_b$ and $J_{bf}$ are the spin-exchange interaction energies. The exchange interaction $J_b<0$ favors a state with fully polarized bosonic spins~\cite{Ho1998,Ohmi1998}. Such a ferromagnetic interaction is considered in this work. Considering a mixture of $^{87}$Rb and $^{6}$Li atoms, the s-wave scattering length for fermions, $^{6}$Li, vanishes due to an accidental cancellation at low magnetic
field~\cite{Houbiers1998,OHara2000}. Thus, $U_f$ is set to 0 in our theory.

\begin{figure}[tpb]
\centering
\includegraphics[angle=0,width=\linewidth]{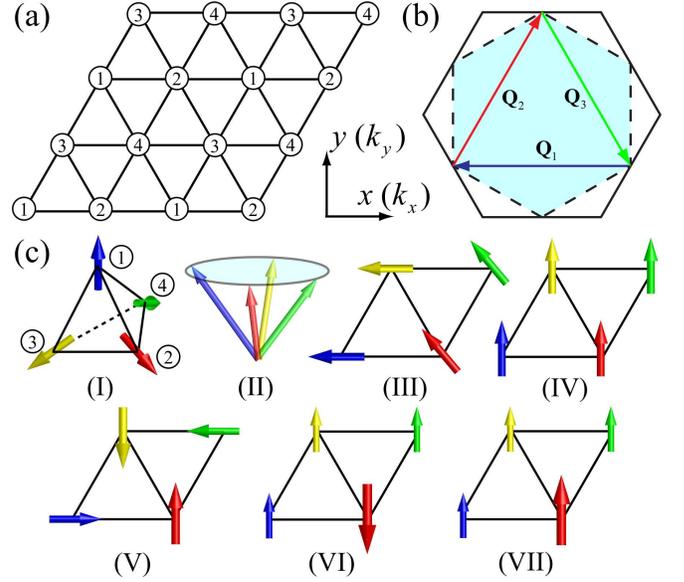}
\caption{(Color online) (a) Schematic picture of a triangular optical lattice and the four-sublattice magnetic ordering. (b) The Brillouin zone of the triangular lattice (solid hexagon) and the Fermi surface (dashed hexagon) at $3/4$ filling. Three nesting wave vectors are $\mathbf{Q}_1=(-2\pi/a,0)$, $\mathbf{Q}_2=(\pi/a,\sqrt{3}\pi/a)$ and $\mathbf{Q}_3=(\pi/a,-\sqrt{3}\pi/a)$, where $a$ is the lattice constant.
(c) The spin configuration of various magnetic orderings of the spin-1 Bose-Einstein condensate, where arrows denote the direction of four spin vectors $\langle\hat{\mathbf{F}}_i\rangle$ ($i=1,2,3,4$): (I) all-out structure chiral spin order; (II) ``umbrella" structure spin order; (III) stripe-cant spin order; (IV) ferromagnetic order; (V) coexistence of DW and $90^\circ$ coplanar spin order; (VI) coexistence of DW and collinear spin order; and (VII) coexistence of DW and ferromagnetic order.
}
\label{fig1}
\end{figure}

{\it Magnetic ordering of bosonic superfluids.}
Here, our theory assumes bosons form a ferromagnetic minicondensate on each lattice site. Such a treatment is well justified when considering the system to be a stack of triangular lattice layers, with a relatively stronger tunneling for bosons than for fermions between nearest layers \cite{suppl}. In the ground state, the system is uniform among different layers. Since bosons locally form a fully polarized ferromagnetic condensation due to interaction effects, the effective model to describe its magnetic moment is a large spin model, in which quantum fluctuations are expected to be suppressed~\cite{Manson1975}. This permits a valid mean-field treatment of bosons.

For spin-$1/2$ fermions, at  $3/4$ filling (meaning $(1/2N_L)\sum_{i}\langle\hat{n}_{f,i}\rangle=3/4$ with $N_L$ is the number of sites), the corresponding Fermi surface as shown in Fig.~\ref{fig1}(b)  is nested, with three nesting vectors
$\mathbf{Q}_1=(-2\pi/a,0)$, $\mathbf{Q}_2=(\pi/a,\sqrt{3}\pi/a)$ and $\mathbf{Q}_3=(\pi/a,-\sqrt{3}\pi/a)$, where $a$ is the lattice constant. Combined with Bose-Fermi density and spin-exchange interactions, the perfect Fermi surface nesting gives rise to ordering instabilities towards formation of commensurate DWs and spin density waves (SDWs) whose unit cells would involve four sublattice sites, shown in Fig.~\ref{fig1} \cite{Martin2008,Akagi2010,Orth2009}.

From the dominant instabilities, we assume in numerics that spin-1 bosons form a magnetic ordering where the density $n_{b,i}=\langle\hat{n}_{b,i}\rangle$ and the spin vector $\langle\hat{\mathbf{F}}_i\rangle$ are periodic under translations of the enlarged unit cell with four sublattices as shown in Fig.~\ref{fig1}(a). Within the superfluid phase,
we take mean-field approximations for bosons, where the bosonic annihilation operator $\hat{b}_{i,\alpha}$ is replaced by its mean value $\phi_{i,\alpha}$. Both the intraspecies ferromagnetic interaction
of bosons and the interspecies spin-exchange interaction
favor a state of boson condensation with spins fully polarized locally.
The corresponding condensate wavefunction is $\phi_i=\sqrt{n_{b,i}}e^{i\xi_i}\zeta_i$, where
\begin{eqnarray}
\zeta_i=\left(e^{-i\varphi_i}\cos^2(\theta_i/2), \sin\theta_i/\sqrt{2},e^{i\varphi_i}\sin^2(\theta_i/2)\right)^T.
\label{orderparameter}
\end{eqnarray}
It yields the spin moment $\langle \hat{\mathbf{F}}_i\rangle=n_{b,i}\mathbf{f}_i$, where $\mathbf{f}_i\equiv(\sin\theta_i\cos\varphi_i, \sin\theta_i\sin\varphi_i,\cos\theta_i)$.
In our calculation, we allow for a finite-momentum condensation by  introducing a parametrization of the phase,
$\xi_i=\mathbf{q}\cdot\mathbf{r}_i+\tilde{\xi}_i$, where nonzero ${\bf q}$ describes the finite momentum and
$\tilde{\xi}_i$ is periodic across different enlarged unit cells.

\begin{figure}[tpb]
\centering
\includegraphics[angle=0,width=3.1in]{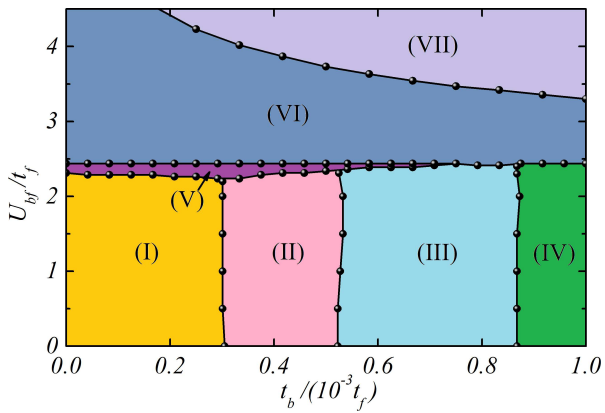}
\caption{(Color online) Zero temperature phase diagram of spinor Bose-Fermi mixtures in a triangular lattice at $3/4$ filling for fermions as functions of $t_{b}$ and $U_{bf}$. Here, $U_b+J_b=3.6\, t_f$ ($U_b>0>J_b$), $J_{bf}=0.4\, t_f$ , $\bar{n}_b=2$ is the averaged number of bosons per site.
}
\label{fig2}
\end{figure}

With the spin configuration of bosons determined by a fixed condensate wavefunction $\phi_i$, the dynamics of fermions is governed by an effective
Hamiltonian:$\hat{H}_f^{\rm eff}=-t_f\sum_{\langle i,j\rangle,\sigma} \hat{c}_{i,\sigma}^{\dagger}\hat{c}_{j,\sigma}+\sum_i[U_{bf}\langle \hat{n}_{b,i}\rangle \hat{n}_{f,i}+J_{bf}\langle\hat{\mathbf{F}}_i\rangle\cdot\hat{\mathbf{S}}_i]$.
The total energy cost is given by
$E[\phi_i] = E_b [\phi_i]+ E_f[\phi_i]$, where $E_b=-t_b\sum_{\langle i,j\rangle,\alpha}\phi_{i,\alpha}^*\phi_{j,\alpha}+\sum_i[U_bn_{b,i}(n_{b,i}-1)+J_b(n_{b,i}^2-2n_{b,i})]/2$
describes the energy of the condensate and $E_f$ is the many-body ground state energy of fermions at $3/4$ filling with respect to $\hat{H}_{f} ^{\rm eff}$ (in our numerics to calculate $E_f$, we discretize the first Brillouin zone into $72\times 72$ points). The variational energy functional $E[\phi_i]$ is then minimized by the simulated annealing method to obtain the ground state.

Fig.~\ref{fig2} summarizes the ground-state phase diagram for a fixed $J_{bf}$ as functions of $t_b$ and $U_{bf}$. The averaged number of bosons per site is chosen as $\bar{n}_b=2$. In the limit of  $U_{bf}=0$ and $t_b=0$, the system is described by a classical Kondo lattice model~\cite{Martin2008,Akagi2010}, where chiral magnetic orders in the ground state are known to occur even with an infinitesimal Kondo coupling $J_{bf} \bar{n}_{b}$ due to Fermi surface nesting. Away from this limit, nonzero boson tunneling and Bose-Fermi density interaction change the ground-state magnetic ordering significantly. The former favors a uniform condensate wavefunction to lower the kinetic energy and will thus suppress spin textures. The latter could induce spatially non-uniform density distributions of bosons. Indeed as boson tunneling is increased, our numerics finds a sequence of states with decreasing spin twists. The corresponding schematic spin configurations are illustrated in Fig.~\ref{fig1}(c). As the Bose-Fermi density interaction is increased, we find that three states denoted by phases (V), (VI), and (VII) are accompanied by DWs.

\begin{figure}[tpb]
\centering
\includegraphics[angle=0,width=\linewidth]{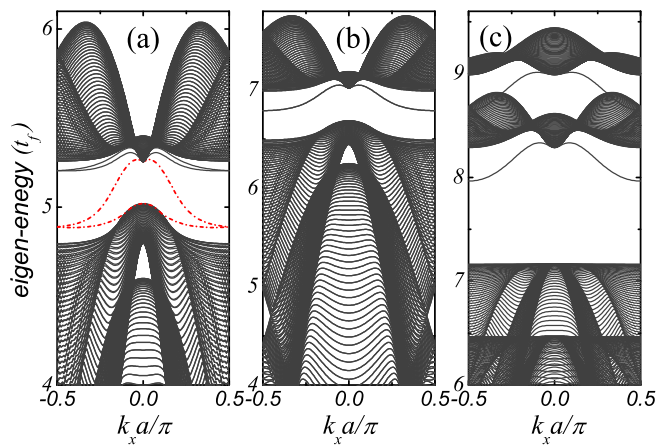}
\caption{(Color online) The spectrum of the Hamiltonian $\hat{H}^{\rm eff}_f$ describing a single fermion moving in a fixed bosonic spin texture background in a strip geometry for different phases at $t_b=0.2\times10^{-3}\, t_f$ and (a) phase (I), $U_{bf}=1.5\, t_f$, (b) phase (V), $U_{bf}=2.35\, t_f$, and (c) phase (VI), $U_{bf}=2.75\, t_f$. Here, red dash-dot lines denote the edge states. }
\label{fig3}
\end{figure}

(1) {\bf Non-coplanar magnetic ordering.}
Numerically, we found two non-coplanar magnetic ordered phases denoted by (I) and (II). They are characterized by a nonzero local spin chirality, which is defined as $\chi_{ijk}=\mathbf{f}_i\cdot\mathbf{f}_j \times\mathbf{f}_k$ on three different lattice sites $i$, $j$, and $k$.

Phase (I) shows the same triple-$\mathbf{Q}$ chiral magnetic ordering found in the Kondo lattice model \cite{Martin2008,Akagi2010}. The boson number density is uniform with $n_{b,i}=\bar{n}_b$ and
\begin{eqnarray}
\mathbf{f}_i=\left(\eta_1 \cos(\mathbf{Q}_1\cdot\mathbf{r}_i),\eta_2\cos(\mathbf{Q}_2\cdot\mathbf{r}_i),
 \eta_3\cos(\mathbf{Q}_3\cdot\mathbf{r}_i)\right),
\label{tripleQ}
\end{eqnarray}
where $|\eta_1|=|\eta_2|=|\eta_3|=1/\sqrt{3}$.
The spin chirality is uniform with  $\chi_{123}=\chi_{243}=\pm 4/(3\sqrt{3})$ which means that fermions experience an effective uniform magnetic flux.

Via coupling to spin textures, fermions are gapped and spontaneously form a quantum Hall insulator leading to quantized Hall conductivity \cite{Martin2008}. To elucidate this behavior, we investigate edge states in a strip geometry, where periodic boundary condition is assumed in the $x$-direction. For a finite length in the $y$-direction, the spectrum for single-particle Hamiltonian $\hat{H}^{\rm eff}_f$ in this strip geometry are plotted in Fig.~\ref{fig3}(a). It clearly shows that there are topological edge states in the gap connecting the upper and lower bulk states.

Phase (II) shows a uniform boson density distribution and a non-coplanar magnetic ordering, where four spin vectors $\langle\hat{\mathbf{F}}_i\rangle$ form a umbrella structure as illustrated in Fig.~\ref{fig1}(c).
The local spin chirality is staggered with a zero summation on the whole lattice and its absolute value is uniform.

(2) {\bf Coplanar magnetic ordering.} Numerically, we find two coplanar phases denoted by (III) and (V).
Within phase (III), bosons are uniformly distributed in every site and four spin vectors form a stripe-cant structure as shown in Fig.~\ref{fig1}(c). Phase (V) shows a $90^{\circ}$ coplanar magnetic ordering accompanied by a DW, where a pair of antiparallel spin vectors have a larger magnitude than another pair of antiparallel spin vectors.

(3) {\bf Collinear magnetic ordering.} The remaining three phases (IV), (VI), and (VII) show collinear magnetic orderings on bosons. More specifically, phase (IV) and phase (VII) demonstrate a ferromagnetic spin ordering, where all spins of bosons are polarized along the same direction. For the phase (VI), in a four-sublattice unit cell three smaller spin vectors with a same magnitude are polarized along opposite direction of the fourth spin vector with a larger magnitude. This magnetic ordering is similar to the SDW of electronic systems driven by thermal fluctuations, which was discussed in Ref.~\cite{Chern2012}.

\begin{figure}[tpb]
\centering
\includegraphics[angle=0,width=\linewidth]{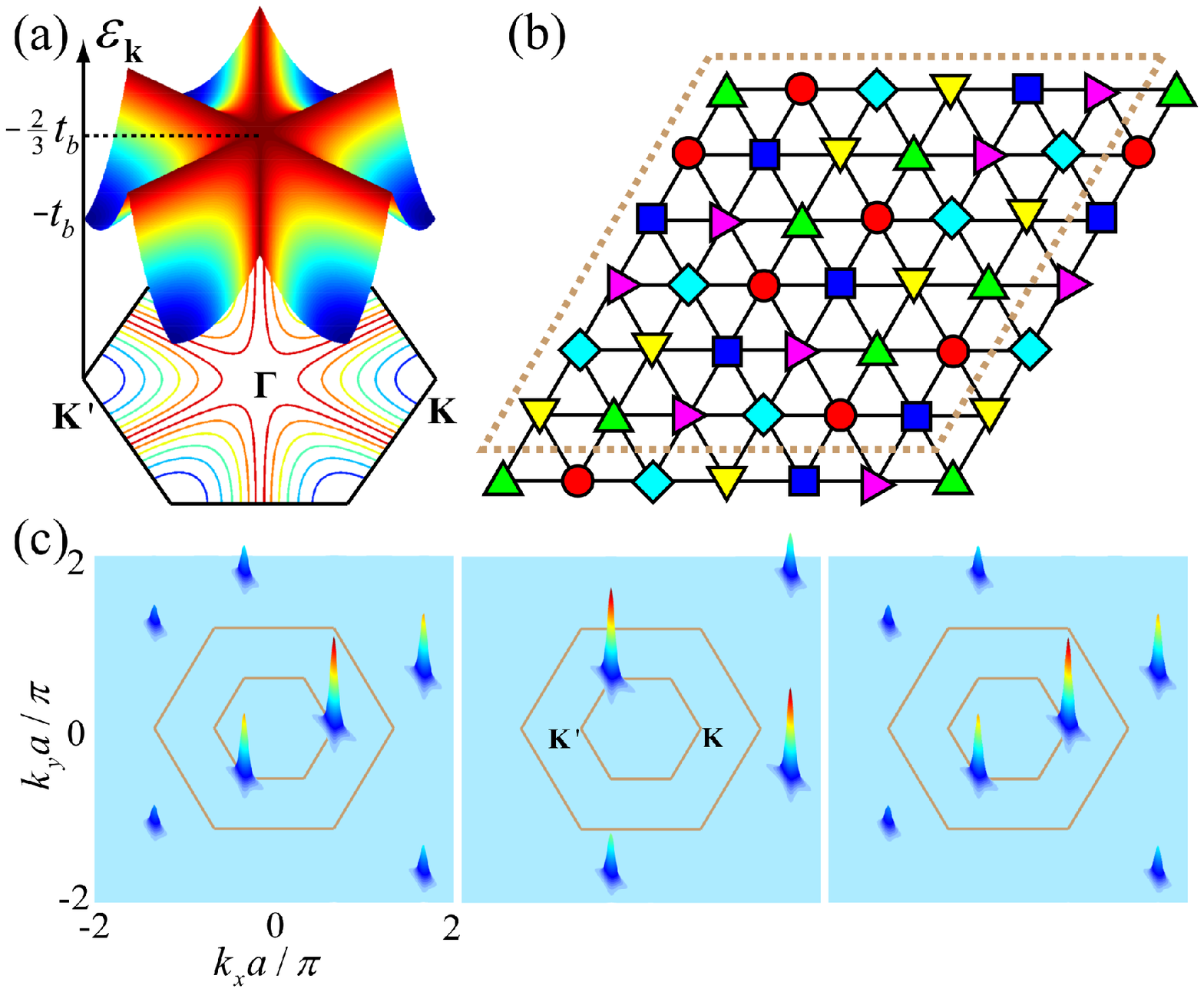}
\caption{(Color online) (a) The lowest energy band of the Hamiltonian $\hat{H}_{b}^{\rm sp}$ with eigen-energy denoted by $\varepsilon_{\mathbf{k}}$. (b) Illustration of the spatially varying phase ($\xi_i$) of condensate wavefunction for the state (I). The symbols, $\square$, $\Scale[\scalefactor]{\circ}$, $\triangle$, $\nabla$, $\Scale[\scalefactor]{\diamond}$ and $\Scale[\scalefactor]{\triangleright}$, represent different phase angles, $0$, $\pi/3$, $2\pi/3$, $\pi$, $4\pi/3$ and $5\pi/3$, respectively. One unit cell of $\xi_i$ is denoted by the dashed rhombus. (c) Time-of-fight pictures for three spin states $M_F=1$ (left), 0 (middle) and $-1$ (right) expected from the experiment for the state shown in (b) as a signal of the finite-momentum-condensate phase. Here, outer hexagon indicates the first Brillouin zone of the triangular lattice and the inner hexagon denotes the first Brillouin zone for the ordered state with an enlarged unit cell with four sublattices. }
\label{fig4}
\end{figure}

{\it Chiral bosonic superfluid.}
Due to the Bose-Fermi interactions, Fermi surface nesting of fermions not only induces spontaneous spin textures of bosons, but also changes the bosonic superfluidity. The reason is that the spin-gauge symmetry of the ferromagnetic spin-1 Bose-Einstein condensate implies that spatial spin configurations can generate mass circulations~\cite{Stamper-Kurn2013,Kawaguchi2012},
which is manifested in the spatially dependent $\xi_i$.
We find that bosons condense at zero momentum with $\mathbf{q}=0$ for phases: (II), (III), (IV), (V), (VI), and (VII). The only exception is found in the phase (I), where bosons condense at a finite momentum with $\mathbf{q}\ne0$. To understand this feature more clearly, we consider an effective Hamiltonian
$\hat{H}_{b}^{\rm sp}=-\sum_{\langle i,j\rangle} \tilde{t}_{b,ij}\hat{d}_i^{\dagger}\hat{d}_j$
for spin-1 bosons with fixed spin directions as that given by Eq.~(\ref{tripleQ}). Here, $\hat{d}_i^{\dagger}$ creates a spin-1 boson at site $i$ with the spin state $\zeta_i$ and $\tilde{t}_{b,ij}=t_b\zeta_i^{\dagger}\zeta_j$ describes the tunneling of bosons in such a spin state. For instance,we
choose $\eta_1=\eta_2=\eta_3=1/\sqrt{3}$, resulting in
$\{\varphi_1=\pi/4,\theta_1=\arccos(1/\sqrt{3})\}$,  $\{\varphi_2=-\pi/4,\theta_2=\pi-\theta_1\}$, $\{\varphi_3=3\pi/4,\theta_3=\pi-\theta_1\}$, and $\{\varphi_4=-3\pi/4,\theta_4=\theta_1\}$. Diagonalizing the Hamiltonian, we find that the lowest energy band shown in Fig.~\ref{fig4}(a) has band minima at $\mathbf{K}=(2\pi/3a,0)$ and $\mathbf{K}'=(-2\pi/3a,0)$. The corresponding eigenstates have equal populations on each site. These two features are underlying reasons for finite-momentum condensation and uniform density distribution of bosons in the phase (I). Take $\mathbf{q}=\mathbf{K}$ as an example, the spatial variation of $\xi_i$ is illustrated in Fig.~\ref{fig4}(b). As $\mathbf{q}$ is nonzero, bosons condense at a finite momentum, thus time-reversal and parity symmetries are broken.

{\it Experimental realization and detection.} The model of Eq.~(\ref{hamiltonian}) can be realized by loading a mixture of spin-$1/2$ $^{6}$Li and spin-$1$ $^{87}$Rb atoms \cite{Silber2005} in the same spin-independent triangular optical lattice. For instance, we choose laser fields at wavelength $\lambda=1064\, \rm nm$. The single photon recoil energy $E_r$ and polarizability $\alpha(\lambda)$ for $^{87}$Rb ($^{6}$Li) atoms are $h\times 2.0$ kHz ($h\times29.2$ kHz) and $689.9\, a_B^3$ ($270.8 \,a_B^3$) where $a_B$ is the Bohr radius and $h$ is the Planck constant ~\cite{De2014,Safronova2006}. Tuning the field intensity, we should be able to create a deep optical lattice for Rb atoms to have small enough $t_b$ comparing to $t_f$ to reach phases (I)-(IV). Changing the laser frequency can further enhance this capability to reverse the relative amplitude of interspecies lattice potential~\cite{Safronova2006}. As in experiments only the interspecies s-wave triplet scattering length ($\sim20\,a_B$) between $^{87}$Rb and $^{6}$Li has been measured \cite{Silber2005}, the parameters in our model, $U_{bf}$ and $J_{bf}$, cannot be determined. Whether Bose-Fermi density interaction are strong enough to induce DWs requires future experimental developments.

In a harmonic trap potential, four phases (I, V, VI, VII), which have a charge gap in fermions, are expected to occupy a finite spatial range as analogous to formation of Mott insulator shells observed in lattice Bose gases. For other phases which do not have a charge gap, they could be in principle sensitive to the trap. To detect various bosonic magnetic ordering, we suggest to use the spin-resolved optical Bragg scattering technique \cite{Corcovilos2010}. The chiral bosonic magnetic ordering can be identified from triple-${\bf Q}$ peaks in spin structure functions. Its phase boundary can be easily probed in time-of-flight measurements because it is the unique state showing non-zero momentum condensation, as illustrated in Fig.~\ref{fig4}(c). Based on previous studies on Kondo lattice models \cite{Kato2010}, we expect that the optimal transition temperature for the chiral magnetic ordering in a Bose-Fermi mixture can reach around $(0.01, 0.1)\, t_f$. Rigorous calculations are left for future study.

{\it Conclusion.} We have studied the ground state of a mixture of spin-$1$ bosons and spin-$1/2$ fermions in a triangular optical lattice. This spinor Bose-Fermi mixture is found to support interesting magnetic orders, such as non-coplanar, coplanar, and collinear spin orders. Most significantly,
there is a triple-$\mathbf{Q}$ chiral magnetic ordered state, featuring a spontaneous quantum Hall effect in fermions and chiral crystalline superfluidity in bosons. It survives at a finite Bose-Fermi density interaction and a very small bosonic tunneling.
The remarkable features we have found for bosons are beyond the physics contained in CKLM, and provide experimental observables for the underlying symmetry breaking in time-of-flight measurements.
Quantum and thermal fluctuation effects on top of the static spin textures could give rise to dynamical gauge fields ~\cite{Banerjee2012,Zohar2012,Tagliacozzo2013}, which is open for future study.

{\it Acknowledgement.}
X.L. would like to thank helpful discussions with A. Rahmani. This work is supported by AFOSR (FA9550-12-1-0079), ARO (W911NF-11-1-0230), DARPA OLE Program through ARO, The Pittsburgh Foundation, and the Charles E. Kaufman Foundation (Z.F.X., W.V.L.). X.L. acknowledges support by JQI-NSF-PFC and ARO-Atomtronics-MURI. Work at Innsbruck is supported by the SFB FOQUS of the Austrian Science Fund, and ERC Synergy Grant UQUAM.

\newpage
\onecolumngrid
\renewcommand\thefigure{S\arabic{figure}}
\setcounter{figure}{0}
\newpage
{
\center \bf \large
Supplemental Material for: \\
Spontaneous quantum Hall effect in an atomic spinor Bose-Fermi mixture\vspace*{0.1cm}\\
\vspace*{0.0cm}
}
\begin{center}
Zhi-Fang Xu$^{1}$, Xiaopeng Li$^{2}$, Peter Zoller$^{3,4}$, and W. Vincent Liu$^{1}$\\
\vspace*{0.15cm}
\small{\textit{$^1$Department of Physics and Astronomy, University of Pittsburgh,
  Pittsburgh, Pennsylvania 15260, USA\\
$^2$Condensed Matter Theory Center and Joint Quantum Institute,\\
Department of Physics, University of Maryland, College Park, MD 20742-4111, USA\\
$^3$Institute for Quantum Optics and Quantum Information of the Austrian Academy of Sciences, A-6020 Innsbruck, Austria\\
$^4$Institute for Theoretical Physics, University of Innsbruck, A-6020 Innsbruck, Austria}}\\
\vspace*{0.25cm}
\end{center}

In this supplementary material, we provide additional
details on (1) optical lattice potentials, (2) physical quantities along two lines: (a) $U_{bf}=1.5\, t_f$ and (b) $t_{b}=0.2\times 10^{-3}\, t_f$, for the phase diagram shown in Fig. 2, and (3) more discussions about  experimental realization and detection.

\section{1. Optical lattice}
To realize the model, we need to create a two-dimensional triangular optical lattice along $x$-$y$ plane and a one-dimensional optical lattice along $z$-axis both for bosons and fermions.

Experimentally, the triangular optical lattice~\cite{Becker2010} can be created by three linear $z$-axis polarized  laser beams that propagate along the $x$-$y$ plane with lattice vectors $\mathbf{k}_1=k_L(1,0,0)$, $\mathbf{k}_2=k_L(-1/2,\sqrt{3}/2,0)$, and $\mathbf{k}_3=k_L(-1/2,-\sqrt{3}/2,0)$, where $k_L$ is the wave vector of the laser beams. Thus, adding up the field strength of each laser beam coherently,  the total field strength is equal to
\begin{eqnarray}
\mathbf{E}_{\rm 2D}(\mathbf{r},t)=\sum\limits_{i=1}^3E_{0}\hat{e}_z \cos(\mathbf{k}_i\cdot\mathbf{r}-\omega_L t+\delta_i).
\end{eqnarray}
Averaging over one period, we obtain a spin-independent potential $\Big(\propto\int_0^{2\pi/\omega_L} |\mathbf{E}_{\rm 2D}|^2d t/(2\pi/\omega_L)\Big)$ given by
\begin{eqnarray}
V_{\rm 2D}(\mathbf{r})=\frac{V_0}{2}\left(\frac{3}{2} +\cos[(\mathbf{k}_1-\mathbf{k}_2)\cdot\mathbf{r}+\delta_1-\delta_2]
+\cos[(\mathbf{k}_2-\mathbf{k}_3)\cdot\mathbf{r}+\delta_2-\delta_3]
+\cos[(\mathbf{k}_3-\mathbf{k}_1)\cdot\mathbf{r}+\delta_3-\delta_1]\right).
\end{eqnarray}
Here, $V_0=4I_0$ and $I_0$ is the light shift produced by one of the lattice beams. The values of $\delta_i$ ($i=1,2,3$) are unimportant for the resulting lattice structure, as they can only globally shift the lattice center.

For a mixture of $^{87}$Rb and $^{6}$Li atoms, if we choose the laser wave length as $\lambda=1064\,\rm nm$, $V_0<0$ for all atoms, and meanwhile the absolute value of $V_0$ for $^{87}$Rb is much larger than that for $^{6}$Li, which enables us to create a deep triangular optical lattice for Rb atoms and simultaneous a shallow triangular optical lattice for Li atoms \cite{Safronova2006s}.

For the one-dimensional optical lattice along $z$-axis, it can be created by imposing two counter propagating laser beams. The resulting lattice potential is given by
\begin{eqnarray}
V_{\rm 1D}(\mathbf{r})=V_z\sin^2(\tilde{k}_Lz).
\end{eqnarray}
According to the polarizability $\alpha(\lambda)$ listed in Ref.~\cite{Safronova2006s}, choosing laser beams with $\tilde{k}_L\ne k_L$, we should be able to create a deep one-dimensional optical lattice for Li atoms to suppress their tunneling between layers and a shallow one for Rb atoms to realize a ferromagnetic condensate.

\begin{figure}[tbp]
\centering
\includegraphics[angle=0,width=4.0in]{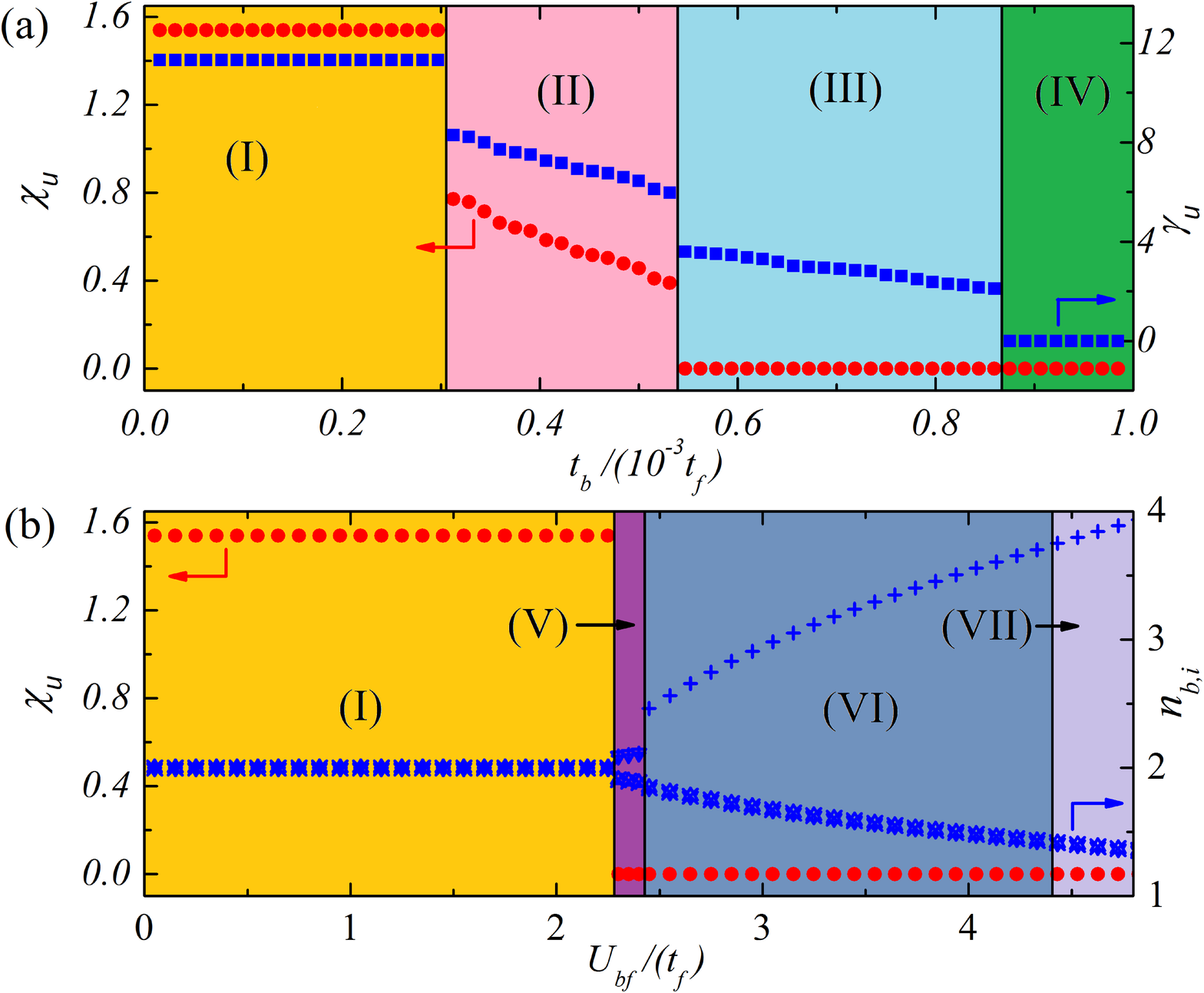}
\caption{(a) Scalar spin chirality $\chi_u\equiv|\chi_{123}|+|\chi_{243}|$ (red filled circles)
and $\gamma_u$ (blue filled squares) defined in a unit cell as a function of
$t_b$ at $U_{bf}=1.5\, t_f$. (b) Local scalar spin chirality $\chi_u$ (red filled circles)
 and the number of bosons $n_{b,i}$ (blue symbols `$\Delta$', `$+$', `$\times$', `$\nabla$') at site $i$ ($i=1,2,3,4$) as a
function of $U_{bf}$ at $t_{b}=0.2\times 10^{-3}\, t_f$.
}
\label{sfig1}
\end{figure}

\section{2. Physical quantities}
Physical quantities used to characterize different phases along two straight lines are shown in Fig.~\ref{sfig1}. Numerically we use $\chi_u\equiv |\chi_{123}|+|\chi_{243}|$ to distinguish non-coplanar phases from others. When bosons form a spontaneous coplanar magnetic ordering, the local scalar spin chirality $\chi_u$ is zero.
We thus define another quantity $\gamma_{ij}=\mathbf{f}_i\times\mathbf{f}_j$,
and use $\gamma_u=\sum_{i\ne j}|\gamma_{ij}|$  with $i,j=1,2,3,4$
to distinguish coplanar and collinear magnetic orderings.

\section{3. Discussions about experimental realization and detection}
In this section, we discuss potential issues of interest to experimental realization and detection.

First, we would like to mention that many parameters in the lattice Bose-Fermi mixture system, such as $t_b$, $U_b$, $J_b$, $U_{bf}$, and $J_{bf}$ are highly tunable simply by controlling the depths of optical lattices, although controllability of the ratios $J_{bf}/U_{bf}$ and $J_b/U_b$ might be limited. From experimental results, we know that the ratio $J_b/U_b$ between spin-exchange and density interactions has orders of magnitude difference for different atoms. For example, it is about $-0.005$ for $^{87}$Rb, $0.04$ for $^{23}$Na and $-0.50$ for $^7$Li~\cite{Stamper-Kurn2013}. For the Bose-Fermi mixture (Rb-Li) we considered, the ratio $J_{bf}/U_{bf}$ between spin-exchange and density interactions is indeed unknown. It may be small but it may be significant as well. However, due to the high tunability of parameters in the lattice Bose-Fermi mixture system, we expect that most of predicted phases are accessible to experiments. It is not possible to predict where the system locates at in the phase diagram.

Second, estimating the transition temperature for this Bose-Fermi mixture system is a very non-trivial question. For the classical Kondo lattice model, previous studies via Monte Carlo simulations find a transition temperature around $0.03 t_f$ for the Kondo coupling chosen to be $2 t_f$ ~\cite{Kato2010}. The Bose-Fermi mixture system contains many more tuning parameters. We therefore expect that the optimal transition temperature may reach $0.1 t_f$, which is then accessible to the present experiments. Rigorous studies of the transition temperature are left for future numerical simulations.

Third, we would like to point out that changing the Fermi filling slightly away from $3/4$ will not change the underlying physics for phases (I), (V), (VI), and (VII) that show a finite energy gap at Fermi surface. The story for other states can be complicated, in principle. Related physics has been investigated for the classical Kondo lattice models (see Ref.~\cite{Akagi2010}), where phases other than a uniform ferromagnet are found away from $3/4$ filling.

\end{document}